\documentclass[review]{article}
\usepackage{pifont}
\usepackage{lineno,hyperref}
\modulolinenumbers[5]

\begin{document}


%

\title{Local SU(3) gauge invariance in Weyl 2-spinor language and quark-gluon plasma equations of motion}

\maketitle

\bigskip


%
\author{J. Buitrago}
\bigskip

{ Department of Astrophysics of the University of La Laguna, Avenida
Francisco Sanchez, s/n, 38205, La Laguna, Tenerife, Spain
and Instituto de Astrofisica de Canarias (IAC), E-38200 La Laguna, Tenerife, Spain}
\bigskip

jgb@iac.es




%


%
%
%

\bigskip

\date{\today}


\begin{abstract}

In a new  Weyl 2-spinor approach to Non abelian Gauge Theories, starting with the local U(1) Gauge group of a previous work, we study now the  SU(3) case corresponding to quarks (antiquarks) interacting with color fields. The principal difference with the conventional approach is that particle-field interactions are not described by means of  potentials but by the field strength magnitudes. Some analytical expressions showing similarities with electrodynamics are obtained. Classical equations that describe the behavior of quarks under gluon fields might be in principle applied to the quark-gluon plasma phase existing during the first instants of the Universe.

\end{abstract} 

%

%

%
\section{Introduction}

In a previous paper \cite{bui} (hereafter Paper I) a {\it classical} $U(1)$ local gauge invariant lagrangian in terms of the electric and magnetic field strengths was introduced  differing to the usual lagrangian quantum approach leading to the Dirac Equation interacting with an external electromagnetic field described by the four potential $A^{\mu}$. Now, in this paper, the mentioned formalism will be  extended to non abelian gauge theories. Given that the Yang-Mills fields, in the SU(2) case, should be massless and are only partially incorporated  in the electroweak theory (devoid of hardly any classical interpretation in the weak sector), it is clear that any classical equations of motion would have only an academic interest. On the other hand, the SU(3) case, involving 8 massless color fields, might be of interest, not only by itself but also in the study of the initial evolution of the Universe (for energies above 100 GeV) when the running coupling constant would be small and thus accesible to a non perturbative study. 

The similitudes between the gluon fields and the ordinary electric and magnetic fields have been noted by many whose names (given their number)  I do not want to remember. In electromagnetism we had, to start with, the Coulomb law (known centuries ago), while for the forces between quarks, at least analytically, there is not something of such a simplicity to start with. It is the main purpose of this work to try to fill this gap and give the quark-color field interaction a classical support. In section 3 we shall see that for colorless (not singlet) color fields it is possible to find, in the special case of a constant electric-like field, the same analytical solution of electrodynamics. For the other color gluon fields, we have not been able to see (or I have not been able to show) whether  any similar result holds, consequently, we are not able to assert that color fields are just replicas of the electromagnetic field. In fact, the non abelian group and the non linear aspects of the interaction seems to point out in the oposite direction.

%
%

As a possible application, at least in principle, the classical equations of motion might be applied to a primordial see of quasi-free quarks and gluons during the first instant of the universe. 



The classical SU(3) local gauge invariant picture that will be presented in the following sections will be symmetrical in both quarks and antiquarks, assuming that the total color charge, and by extension total electric charge, in the early universe was zero (The reader should be warned about the possibility that in a high energy scenery in which both electromagnetic and strong running coupling constant could  eventually be comparable the quark-gluon dynamics description might be in practice incomplete).

\section{The SU(3) classical local gauge invariant lagrangian density}
As in some parts of what next follows we shall implicitly deal with an scenario in which asymptotically free quarks and antiquarks coexist, it will be convenient to change the notation used in Paper I where complex conjugate spinors, such as $\bar\eta^{A'}$, are denoted with a bar over  Greek and Roman symbols and primed capital letters as superscript and subscripts. The {\it bar-over} is conventionally used for antiparticles and we shall obey  this rule naming complex conjugate spinors, of first and higher rank, simply $\eta^{A'}, \varphi^{A'B'}....$ and so on (a discussion about the various conventions used in the 2-Spinor formalism can be found in the book of Penrose and Rindler {\it opus cit.}\cite{penrose}).

 Assuming, for simplicity, a single flavour of quark, which can exist in three color-charge states,
the four momentum can be written in the same way that was done in Paper I, namely 
\begin{equation} \label{momentum}
p^{AA'}_c=\frac{1}{\sqrt{2}}\left[\pi^{A}\pi^{A'}+\eta^{A}\eta^{A'}\right]_c,
\end{equation} \\
while for an antiquark
\begin{equation} \label{bmomentum}
p^{AA'}_c=\frac{1}{\sqrt{2}}\left[\bar\pi^{A}\bar\pi^{A'}+\bar\eta^{A}\bar\eta^{A'}\right]_c,
\end{equation} \\
where $c$ can be any of $1=red, 2=green, 3=blue $. 
Since $p^{AA'}_c$ is to represent the four-momentum of a massive particle, must be time-like and, for any of the three possible values of $c$, fulfill the condition:
\begin{equation} \label{condition}
p^{AA'} p_{AA'} =m^2.
\end{equation} \\

For a free quark (or antiquark putting bars over the symbols) in one of the three colors (anticolors), the classical lagrangian density (with dimensions energy per unit length) would be

\begin{equation} \label{1.2}
{\cal L}_f=\pi_{Ac}\dot{\eta}^{Ac},
\end{equation} 
(dot denoting derivative respect to proper time $\tau$).
As $\eta^{Ac}$ and $\pi_{Ac}$ are triplets, the notation should again be changed relative to Paper I to assure that every term in the lagrangian is a scalar. The $\eta$ and $\pi$ spinors will be represented by
\begin{equation}
\left(\begin{array}{c} \eta^{A1} \\ \eta^{A2} \\ \eta^{A3} \end{array}\right),
\end{equation}
and
\begin{equation}
\left( \pi_{A1} \\ \pi_{A2} \\ \pi_{A3} \right).
\end{equation}
The Euler-Lagrange equations associated to $\eta^{Ac}$ are \\

\begin{equation} \label{1.3}
\frac{d}{d\tau}\frac{\partial \cal L}{\partial \dot{\eta}^{Ac}}-\frac{\partial \cal L}{\partial \eta^{Ac}}=0,
\end{equation} \\
 leading to \\
\begin{equation} \label{1.5}
\dot{\pi}_{Ac}=0 \Longrightarrow \pi_{Ac}=const. \\
\end{equation} \\
The free lagrangian is not invariant under $SU(3)$ local phase transformations 
\begin{equation} \label{trans}
\eta^{Ac}\rightarrow \eta^{'Ac}=exp \left (i \frac{g_s}{2} \lambda_q.\alpha(\tau)^q \right )\eta^{Ac},
\end{equation} \\
and
\begin{equation} \label{trans}
\pi_
{Ac}\rightarrow \pi_{Ac}'=exp \left (- i\frac{g_s}{2} \lambda_q. \alpha(\tau)^q \right )\pi_{Ac},
\end{equation} \\
where $\lambda_q$ ($q=1,2,...,8$) are the Gell-Mann $SU(3)$ matrices.

To restore invariance, the free lagrangian need to acquire an additional term. The procedure similar to the conventional in non abelian gauge theories is to replace the ordinary derivative by the covariant one and consider an infinitesimal transformation. In our classical approach the covariant derivative is defined trough the condition of transforming in the same way as the spinors:
\begin{displaymath} \label{covariant}
\frac{D}{d\tau}\eta^{Ac} \rightarrow  exp \left (i \frac{g_s}{2} \lambda_q.\alpha(\tau)^q \right )\frac{D}{d\tau}\eta^{Ac}.
\end{displaymath} \\
After some lengthy algebra, it is found that the interacting gauge invariant lagrangian is

\begin{equation}\label{int}
{\cal L}_f=\pi_{Ac}\dot{\eta}^{Ac} - \frac{g_s}{m} \pi_{Ba}\left [\lambda_q {W_A}^{Bq}\right]_{ab} \eta^{Ab}.
\end{equation}

The eight field strengths ${W_A}^{Bq}$ are related to the usual symmetric $W_{AB}^i$ following the rule $W_{AB}^i = \epsilon_{BC} W_A^{Ci}$ while the gauge fields $W^{i}_{AB}$ transform as ($i,j,k=1,2,...,8$) \\

\begin{displaymath}
W^i_{AB} \rightarrow W^i_{AB}-im\dot\alpha^i\epsilon_{AB}-f^{ijk}\alpha^j(\tau)W^k_{AB}
\end{displaymath} 
\begin{displaymath}
{W}^{i}_{A'B'} \rightarrow {W}^{i}_{A'B'}+im\dot\alpha^{i}\epsilon_{A'B'}-f^{ijk}\alpha^{j}
(\tau) {W}^{k}_{A'B'},
\end{displaymath} \\
being $f^{ijk}$ the structure constants of $SU(3)$. The corresponding eight fourth-rank  antisymmetric spinor fields are represented, in the conventional form, as \\
\begin{equation} \label{covariant12}
G^{i}_{AA'BB'}=W^{i}_{AB}\epsilon_{A'B'}+{W}^{i}_{A'B'}\epsilon_{AB}
\end{equation} \\
and, as consequence of the non abelian field, $G^i_{AA'BB'}$ is not invariant but transforms as \\
\begin{displaymath}
G^{i}_{AA'BB'} \rightarrow G^{i}_{AA'BB'}-g_s \left[f^{ijk}\alpha^{j}(\tau)G^{k}_{AA'BB'}\right].
\end{displaymath} \\

with the structure constant of the group:

$$ f^{123} = 1, f^{147} = f^{243} = f^{257} = f^{345} = f^{516} = 1/2, \  f^{458} = f^{678} = \sqrt 3/2.
$$
The Equations of motion can be obtained via the Hamiltonian density
\begin{equation}
{\cal H} = \frac{\partial {\cal L}}{\partial \dot\eta^{Aa}}\dot\eta^{Aa} - \cal L,
\end{equation} \\
being just the interaction term:
\begin{equation}
{\cal H} = \frac{g}{m} \pi_{Ba}\left [\lambda_q {W_A}^{Bq}\right]_{ab} \eta^{Ab}.
\end{equation}
Using Hamilton Equations
\begin{equation}
\dot\pi_{Aa} = - \frac{\partial \cal H}{\partial\eta^{Aa}}
\end{equation}
\begin{equation}
\dot\eta^{Aa} = \frac{\partial \cal H}{\partial\pi_{Aa}}.
\end{equation}
The Equations of motion are finally 
\begin{equation}
\dot\eta_A^a = \frac{g}{m}\left[\lambda_q {W^q_A}^B\right]_{ab}\eta_B^b
\end{equation}
\begin{equation}
\dot\pi_{Aa} = - \frac{g}{m}\pi_{Ba}\left[\lambda_q {W^q_A}^B\right]_{ab}.
\end{equation} 
The goal is then to solve the previous equations in a given environment or background field given by the matrix on the right side and obtain the evolution of the four-momentum (\ref{momentum}) as a function of proper time. Note that as a classical equation there is no distinction between quarks and antiquarks.

\section{Classical picture of color fields and analogies with the electromagnetic field}

In the standard version of QCD, the interaction is described by eight color potentials ${B^l}_\mu$ related to the field strengths by
\begin{equation}\label{gb}
{G^l}_{\mu\nu} = \partial_\nu {B^l}_\mu - \partial_\mu {B^l}_\nu + ig_sf^{jkl} {B^j}_\mu {B^k}_\nu
\end{equation}. \\

Since spacetime derivatives do not change color, in the high energy regime corresponding to small running coupling constant $g_s$, the color fields seems to be just a copy of the electromagnetic field and, therefore, in the early almost homogeneous universe, we can expect that three-gluon vertex (proportional to $g_s$) as well as four-gluon vertex (proportional to $g_s^2$) are almost absent .

Each of the massless (spin one) color field potentials ${B^l}_\mu$ has two helicity states and they dress themselves in eight different color combinations corresponding to the color octet \cite{griff}:
\begin{equation} \label{assignment}
\begin{array}{c} B^1 \rightarrow (r\bar b + b\bar r)/ {\sqrt 2} \\ B^2 \rightarrow  -i(r\bar b - b\bar r)/{\sqrt 2} \\ B^3 \rightarrow  (r\bar r - b\bar b)/{\sqrt 2} \\ B^4  \rightarrow (r\bar g + g\bar r)/{\sqrt 2}  \\ B^5 \rightarrow -i(r\bar g - g\bar r)/{\sqrt 2} \\ B^6 \rightarrow (b\bar g + g\bar b)/{\sqrt 2} \\ B^7 \rightarrow -i(b\bar g - g\bar b)/{\sqrt 2} \\ B^8 \rightarrow (r\bar r + g\bar g - 2 b\bar b)/{\sqrt 6}.
 \end{array}
\end{equation}
Two of them, $B^3$ and $B^8$, are colorless while the rest mix color charges. In our description, this will be reflected in the diagonal and non diagonal terms in the equations below.
In a homogeneous early universe the background gluon field should have been homogeneous and isotropic enough, otherwise the universe could possibly not have evolved to the state revealed by the 3K radiation field.

Once expanded, the classical Equations of Motion looks like
\begin{equation} \label{expequations}
\left(\begin{array}{c}
\dot{\eta}^{1A} \\
\dot{\eta}^{2A} \\
\dot{\eta}^{3A} \\
\end{array}\right)=\frac{g_{s}}{m}\left(\begin{array}{ccc}
W^{3B}_{A}+\frac{W^{8B}_{A}}{\sqrt{3}} & W^{1B}_{A}+iW^{2B}_{A} & 
W^{4B}_{A}+iW^{5B}_{A} \\
W^{1B}_{A}-iW^{2B}_{A} & -W^{3B}_{A}+\frac{W^{8B}_{A}}{\sqrt{3}} &
W^{6B}_{A}+iW^{7B}_{A} \\
W^{4B}_{A}-iW^{5B}_{A} & W^{6B}_{A}-iW^{7B}_{A} &
-2\frac{W^{8B}_{A}}{\sqrt{3}} \\
\end{array}\right)
\left(\begin{array}{c}
\eta^{1B} \\
\eta^{2B} \\
\eta^{3B} \\
\end{array}\right)
\end{equation} \\
\begin{equation} \label{eqmot32} 
\left( 
\dot{\pi}_{1A} \\
\dot{\pi}_{2A} \\
\dot{\pi}_{3A} \\
\right)=-\frac{g_{s}}{m}\left({\pi}_{1B} \\
{\pi}_{2B} \\
{\pi}_{3B} \\
\right) 
\left(\begin{array}{ccc}
W^{3B}_{A}+\frac{W^{8B}_{A}}{\sqrt{3}} & W^{1B}_{A}+iW^{2B}_{A} & 
W^{4B}_{A}+iW^{5B}_{A} \\
W^{1B}_{A}-iW^{2B}_{A} & -W^{3B}_{A}+\frac{W^{8B}_{A}}{\sqrt{3}} &
W^{6B}_{A}+iW^{7B}_{A} \\
W^{4B}_{A}-iW^{5B}_{A} & W^{6B}_{A}-iW^{7B}_{A} &
-2\frac{W^{8B}_{A}}{\sqrt{3}} \\
\end{array}\right).
\end{equation}\\
In the precedent equations, the diagonal terms do not change the quark color while the non diagonal connect different colors. However, since color must be conserved in any of the  resulting differential equations, it is convenient to check further this point and also see if there is agreement between the field strengths distribution in (\ref {expequations}) and the color assignment to the different quark and antiquark wave functions that describe interactions at the quantum level. In our approach, embedded in the eight field combinations  are color quanta which should be classified according to the assignment (\ref{assignment}). First let us see the diagonal terms:
\begin{equation}\label{diagonal}
W^{3B}_{A}+\frac{W^{8B}_{A}}{\sqrt{3}} \rightarrow  \frac{(r\bar r - b\bar b)} {\sqrt 2} + \frac {(r\bar r + b\bar b - 2g\bar g)} { \sqrt 3 \sqrt 6} = \frac {1}{\sqrt2} \left (\frac{4}{3}r\bar r - \frac {2}{3}b\bar b - \frac {2}{3}g\bar g \right ).
\end{equation}
In the same way
\begin{equation}
-W^{3B}_{A}+\frac{W^{8B}_{A}}{\sqrt{3}} \rightarrow \frac{1}{\sqrt 2}\left( \frac{4}{3}b\bar b -\frac {2}{3}r\bar r - \frac {2}{3}g\bar g \right )
\end{equation}
\begin{equation}
-\frac{2W^{8B}_{A}}{\sqrt{3}} \rightarrow \frac {1}{\sqrt 2}\left( \frac{4}{3}g\bar g - \frac{2}{3} r\bar r - \frac{2}{3} b\bar b \right ).
\end{equation}
As should be expected, all diagonal terms have a similar status thus respecting $SU(3)$ symmetry.
For the non diagonal terms we have the following association:
\begin{equation} \label{nondiagonal}
W^{1B}_A + i W^{2B}_A \rightarrow \frac{r\bar b + \bar r b}{\sqrt 2} + \frac{r\bar b - b\bar r}{\sqrt 2} = \frac {2}{\sqrt 2} r\bar b
\end{equation}
\begin{equation} \label{non2}
W^{1B}_A - i W^{2B}_A \rightarrow   \frac {2}{\sqrt 2} b\bar r
\end{equation}
\begin{equation} \label{non3}
W^{4B}_A + i W^{5B}_A \rightarrow \frac{2}{\sqrt 2} r\bar g,
\end{equation}
\begin{equation} \label{non4}
W^{4B}_A - i W^{5B}_A \rightarrow \frac{2}{\sqrt 2} g\bar r,
\end{equation}
\begin{equation} \label{non5}
W^{6B}_A + i W^{7B}_A \rightarrow \frac{2}{\sqrt 2} b\bar g,
\end{equation}
\begin{equation} \label{non6}
W^{6B}_A - i W^{7B}_A \rightarrow \frac{2}{\sqrt 2} g\bar b.
\end{equation}
%


%
We cannot forget that our equations are classical. A quark of a certain color can interact with a gluon and change its color but this kind of interaction, described in QCD, is difficult to describe in classical lenguaje. If we can obtain some information about the nature of color forces, we have to get some information about how does a quark move under a certain color field. To this end we shall first assume that there is a small lapse of time in which the particle moves under an specific color field and to simplify even more, a colorless constant gluon field. The color gluon field case will be discussed afterwards.

Obviously, in the high energy regime, and in the conventional approach, once the nonlinear terms in the potentials $B^i_\mu$ are removed, the similarities with the electromagnetic field are  complete. In what follows and with the double purpose of showing in some detail how to obtain $\vec p(t)$ from the spinor equations and and to see whether the solution coincides with the electromagnetic one, they will be solved  in the simple case of a constant electric-like field in some direction. In the electromagnetic case we know the solution:
$$ 
\vec p(t) = \vec E t.
$$
Assume that we have an isolated quark of color 3 in a background of electric-like $\cal E$ and magnetic-like $\cal B$ color fields (see equation (50) in Paper I). In such a case, the expression for the field would be

\begin{equation}
W ^{8A}_{B}= \frac {1}{2}  
\left[\begin{array}{cc} {\cal E}_3& {\cal E}_{1}+i {\cal E}_{2} \\ {\cal E}_{1}-i {\cal E}_{2}& -{\cal E}_{3} \end{array}\right] + i \frac{1}{2}
\left[\begin{array}{cc} {\cal B}_{3} & {\cal B}_{1}+i {\cal B}_{2} \\ {\cal B}_{1} - i {\cal B}_{2} & - {\cal B}_{3} \end{array}\right].
\end{equation}
If for simplicity the field is pure electric-like, constant and in the $z$ direction:
\begin{equation}
W^{8A}_B = \left(\begin{array}{cc} \cal E & 0 \\ 0 & -\cal E \end{array}\right).
\end{equation}
The equations to solve are
\begin{equation}
\dot\eta^{A3} = \frac{g_s}{2m}\left (-\frac{2}{\sqrt 3}W^{8A}_B \right)\eta^{B3}
\end{equation}
and
\begin{equation}
\dot\pi_{A3} = - \frac{g_s}{2m}\pi_{B3}\left (-\frac{2}{\sqrt 3} W^{8B}_A \right)
\end{equation}
The solution in terms of the {\it physical} contravariant spinor components is given by
\begin{equation}
\eta^{03} = \sqrt m \exp \left ( -\frac {g_s}{m\sqrt 3}{\cal E} \tau \right )
\end{equation}
\begin{equation}
\eta^{13} = \sqrt m \exp \left ( \frac {g_s}{m\sqrt 3}{\cal E} \tau \right ),
\end{equation}
with equal solution for $\pi^{A3}$ (it has been assumed that at proper time $\tau =0$ the particle is at rest thereby the factor $\sqrt m$). Now, from the spinor transcription of the four momentum given by equation (1) it is  relatively easy to find out the expressions for the energy and spatial components as function of proper time:
\begin{equation}
p^0 = E = \frac{1}{2\sqrt 2}m \cosh \left (\frac{2g_s}{m\sqrt 3}{\cal E}\tau \right )
\end{equation}
\begin{equation}
p^3 = \frac{1}{2\sqrt 2}m \sinh \left (\frac{2g_s}{m\sqrt 3}{\cal E}\tau \right ).
\end{equation}
$$ p^1 = p^2 = 0 $$
Finally from the relationship between laboratory time and proper time when the driving force is constant
\begin{equation}
t = \frac{1}{C} \sinh (C \tau),
\end{equation} 
\begin{equation}
p^3 = \frac{1}{\sqrt 2}  \left (\frac{g_s}{\sqrt 3}{\cal E}\right )t.
\end{equation}
As in electrodynamics, this is an exact solution for a test particle (quark) ignoring radiation effects and under the special environment considered. We would like to point out that equations (\ref{expequations}) and the next,  being the result of local gauge invariance, are exact. To clarify further this point note that the spinor field strength $G^i_{AA'BB'}$ is just the translation to the 2-spinor lenguaje of the field tensor $G^i_{\mu \nu}$, in the particular case of $G^8_{\mu \nu}$, the non linear term is
\begin{equation}
{G^8_{\mu\nu}}_{nl} = ig_s f^{ij8}B^j_\mu B^k_\nu,
\end{equation}
with $f^{458} = f^{678}= \sqrt 3 /2$. To show explicitly that this term is included in $G^8_{AA'BB'}$, one only needs  to rewrite Eq (\ref{covariant12})  in terms of the potentials instead of the field strengths.

From the similitude found in equations (23) to (25) for the three colors, it is apparent that the result found for a quark in color 3 would also be the same for the other two (as required from the $SU(3)$ symmetry). 

Before entering in a few calculations related to the non diagonal elements in (\ref{nondiagonal}) that incessantly change the quarks colors, it seems in place a few comments on the difference between the color fields associated to diagonal entries and the rest of them. While the diagonal fields, that do not change the quark color charge, fit in the classical picture of {\it something} laying in the background that silently act on the particles. The non diagonal fields do not fit into this kind of picture, instead, its representation or whatever one might call, is more adequate inside the QCD theory describing fields as some kind of quanta interacting with the quarks in the particle sense. In plain words, a sea of color quanta in interaction with another sea of quarks and antiquarks. As we shall see, in such circumstances, our classical equations enter in some trouble.

In contrast to the diagonal terms,
for the non diagonal entries in (\ref{expequations}) we have the following set of three differential equations system, connecting two different color charges:

\begin{equation}\label{sys1}
\begin{array}{c}
\dot\eta^{A1} = \frac{g_s}{m}(W^{1A}_B + iW^{2A}_B)\eta^{B2} \\
\dot\eta^{A2} = \frac{g_s}{m}(W^{1A}_B - iW^{2A}_B)\eta^{B1} \\
\end{array}
\end{equation} \\
\begin{equation}\label{sys2}
\begin{array}{c}
\dot\eta^{A1} = \frac{g_s}{m}(W^{4A}_B + iW^{5A}_B)\eta^{B3} \\
\dot\eta^{A3} = \frac{g_s}{m}(W^{4A}_B - iW^{5A}_B)\eta^{B1} \\
\end{array}
\end{equation} \\
\begin{equation}\label{sys3}
\begin{array}{c}
\dot\eta^{A2} = \frac{g_s}{m}(W^{6A}_B + iW^{7A}_B)\eta^{B3} \\
\dot\eta^{A3} = \frac{g_s}{m}(W^{6A}_B - iW^{7A}_B)\eta^{B2} \\
\end{array}
\end{equation}
Comparing with the gluon-color content of the six relations (26) to (31), we see that the quark color changes at a quark-gluon vertex in a way consistent with the quantum description (see [3] Chapter 9). 
Let us now choose one of the system, for instance the second one. By derivation respect to proper time and considering both fields as constant:
\begin{equation}
\frac {d^2}{d \tau^2} \eta^{A1}= \frac{g_s^2}{m^2} (W^{4A}_B + iW^{5A}_B)(W^{4A}_B - iW^{5A}_B) \eta^{B1}.
\end{equation}
To avoid the complications of the non commutative matrix algebra, both fields will be chosen as pure electric-like and in the z-direction. In such a case, the equation to solve is
\begin{equation}
\frac {d^2}{d \tau^2} \eta^{A1} =  \frac{g_s^2}{m^2} W^A_B \eta^{B1},
\end{equation}
with
\begin{equation}
W^A_B = \left(\begin{array}{cc} {\cal E}^{2}_1+ {\cal E}^2_2 & 0 \\ 0 &  {\cal E}^2_1+ {\cal 
E}^2_2 \end{array}\right).
\end{equation}
In the rest frame of the red quark, the solution is the same for both components of the spinor, omitting the color index for clarity
\begin{equation}
\eta^{0,1} = + \sqrt m \cosh (k\tau),
\end{equation}
where 

$$ 
k= \frac{g_s}{m}\sqrt{{\cal E}^2_1 + {{\cal E}^2_2}}.
$$
The solution for $\pi^A$ is similar only with a minus sign for the $\pi^1$ component 
\begin{equation}
\pi^{0,1} = \pm \sqrt m \cosh (k\tau).
\end{equation}
Now from Eq.(1) for $p^{AA'}$ we find the interesting but unphysical result for a $m \ne 0 $ particle:
\begin{equation}
p^0 = E = p^3 = \frac{m}{\sqrt 2} \cosh ^2 (k\tau),
\end{equation}
and $p^1 = p^2 = 0$.

We know that conservation of energy and momentum forbids an electron to absorb, or emit, one photon (in its rest system or in any other) and the same should apply  to quarks and gluons. Looking at the initial system of equations and the colors associated to the intervening fields in the process, from the quantum perspective, the process could be described as: "a red quark absorbs a $g\bar r$ gluon becoming a green quark and shortly afterwards (to allow permission from the Heisenberg principle) emits a $r\bar g$ gluon becoming again a red quark". Notwithstanding that the sequence of color changes (see (\ref{non4}) and (\ref{non3})) is coherent in the equations involved, it is clear that the classical (continuous) description is in those cases, at least, inadequate, thereby my prior comments about the somewhat failure of the field classical description for the colored (non diagonal) gluon fields.
\section{Conclusion}
As mentioned at the Introduction, It was my purpose to explore the classical aspects of QCD and obtain some results that might be of relevance. In fact there is a similitude between the Lorentz Force of electrodynamics and its equivalent in {\it chromodynamics}. In Eq. (1), we find  the translation to spinor form ($p^{AA'}$) of the four momentum $p^\mu$ which follows from the isomorphism between real four-vectors and hermitian spinors of second rank. From (12) we could build the chromodynamic "Lorentz Force" analogue as
\begin{equation}
\frac{d p^{AA'}}{d\tau} = G_i^{AA'BB'}p_{BB'}.
\end{equation}
However, in this form many (sometimes non classical) aspects revealed in the spinor equations (21)-(22) would be hidden (see Paper I). In the diagonal terms of the mentioned spinor equations of motion, we have found a certain paralelism with electrodynamics while in the equations related to non diagonal terms there are aspects either paradoxical or, perhaps, unveiling properties of color dynamics that should be worth of further study.
%

%

%


\begin{thebibliography}{99}

\bibitem{bui} Buitrago J. Results in Physics 6 (2016) 346-351
%
%
\bibitem{penrose} R. Penrose and W. Rindler \textit{Spinors and Space-Time}, 
Cambridge Monographs in Mathematical Physics, Vol. $1$, Cambridge Universtiy Press, Cambridge, 
England ($1984$/$1986$).
%
\bibitem{griff} D. Griffiths \textit{Introduction to Elementary Particles} John Wiley and Sons, Inc ($1987$)
%
\end{thebibliography}
\end{document}